# DISSIPATIVE SYSTEMS ENTROPY

**E.E. Perepelkin[a], B.I. Sadovnikov[a], N.G. Inozemtseva[b]**

[a] *Faculty of Physics, Lomonosov Moscow State University, Moscow, 119991 Russia*
E. Perepelkin e-mail: pevgeny@jinr.ru, B. Sadovnikov sadovnikov@phys.msu.ru
[b] *Dubna State University, Moscow region, Moscow, 141980 Russia*
e-mail: nginozv@mail.ru

**Abstract**

In this paper, we introduce the generalized phase space $\left(\vec{r},\vec{v},\dot{\vec{v}},\ddot{\vec{v}},...\right)$, which expands the known phase space $\left(\vec{r},\vec{v}\right)$. The fact is that the introduced space is the infinity dimensional phase space.

The paper shows that dissipative systems in generalized phase space may be considered as conservative systems.

It is shown that, in infinity dimensional phase space, the entropy is a constant value. It is shown that the transition to finite dimensional phase space leads to dissipations and change of the entropy.

The paper contains the rigorous mathematical result.

## 1. Introduction

The construction of rigorous mathematical methods to describe physical phenomena is one of the main problems of theoretical and mathematical physics [1-4]. The phase space and Liouville's theorem [5] on conservation of the phase space $\left(\vec{r},\vec{v}\right)$ (where $\vec{r}$ is coordinate and $\vec{v}$ is velocity) volume underlie the classical mechanics, statistical physics and quantum mechanics.

Considering dissipative systems in statistical physics [6], plasma physics [7], accelerator physics, simulating controlled nuclear fusion there may occur situations when phase volume is not reserved in space $\left(\vec{r},\vec{v}\right)$.

In 1850, Ostrogradsky developed high order mechanics [8]. High order mechanics considers Lagrange function $L$ dependent not only on coordinate $\vec{r}$ and velocity $\vec{v}$, but also on acceleration $\dot{\vec{v}}$, and high order accelerations $\ddot{\vec{v}},\dddot{\vec{v}},...,\overset{(n)}{\vec{v}}$, that is $L=L\left(\vec{r},\vec{v},\dot{\vec{v}},\ddot{\vec{v}},...\overset{(n)}{\vec{v}};t\right)$ [30,31].

The quantum field theory, the gauge theories [9-11], general relativity [12-17] and some string models [18, 19] use the high order mechanics. In the middle of the 20th century, Vlasov considered in [20-23] the similar method for the probability distribution density functions $f\left(\vec{r},\vec{v},\dot{\vec{v}},\ddot{\vec{v}},...\overset{(n)}{\vec{v}};t\right)$.

By analogy with the classic phase space $\left(\vec{r},\vec{v}\right)$, it is seems natural to introduce infinity generalized phase space $\left(\vec{r},\vec{v},\dot{\vec{v}},\ddot{\vec{v}},...\right)$ into problems in high order mechanics. Thus, a question of an analogue to Liouville's theorem and Liouville's equation for generalized phase space $\left(\vec{r},\vec{v},\dot{\vec{v}},\ddot{\vec{v}},...\right)$ comes up.

*The paper is aimed at generalization of Liouville's equation and Liouville's theorem on conservation of phase space volume $(\vec{r},\vec{v})$ in case of the infinity dimensional generalized phase space $(\vec{r},\vec{v},\dot{\vec{v}},\ddot{\vec{v}},...)$.*

The practical implications of such generalization is that it makes it possible to describe dissipative systems as conservative ones. That is a dissipative system in the generalized phase space $(\vec{r},\vec{v},\dot{\vec{v}},\ddot{\vec{v}},...)$ may be considered as a conservative one. Thus, the spectrum of analytical and numerical methods for the description and investigation of dissipative systems is expanded essentially.

The main definitions and theorems are listed in §1, 3 of the paper. The proofs of the theorems are put in Appendix for convenience. An example of the calculation of the phase volume in space $(\vec{r},\vec{v})$ and in space $(\vec{r},\vec{v},\dot{\vec{v}})$ for a multi-particle system is contained in §2. It is shown that in spaces $(\vec{r},\vec{v},\dot{\vec{v}})$, $(\vec{v},\dot{\vec{v}})$ and $(\dot{\vec{v}})$, the phase volume is conserved, and in space $(\vec{r},\vec{v})$ the phase volume is not conserved. The obtained results are in full concordance with the theorem statements from §1.

In §3, a generalized Liouville's equation for infinity dimensional phase space $(\vec{r},\vec{v},\dot{\vec{v}},\ddot{\vec{v}},...)$ is obtained. It is shown that the phase «liquid» in space $(\vec{r},\vec{v},\dot{\vec{v}},\ddot{\vec{v}},...)$ is incompressible, which leads to the entropy constancy. The transition to finite dimensional phase spaces leads to production of the entropy sources, to compressibility of the phase «liquid», to the presence of dissipative processes.

Generalized Liouville's equation at the transition to finite dimensional phase spaces is divided into an infinite chain of the Vlasov equations for the probability density distribution functions $f_1(\vec{r},t), f_2(\vec{r},\vec{v},t), f_3(\vec{r},\vec{v},\dot{\vec{v}},t),....$. The modified Vlasov equation [6] is transformed into classical Liouville's equation in case no dissipations occur.

Conclusion can be found at the end of the paper, which contains the main results of the work.

## §1 Generalized Liouville's theorem

***Definition 1.*** *Let $^{\Psi}\Omega$ be an infinite-dimensional space with elements $\vec{\xi} \in {^{\Psi}\Omega}$, where*

$$\vec{\xi} = \{\vec{r},\vec{v},\dot{\vec{v}},\ddot{\vec{v}},...\}^T \in {^{\Psi}\Omega} \text{ or } \vec{\xi} = \{x,y,z,v^x,v^y,v^z,\dot{v}^x,\dot{v}^y,\dot{v}^z,...\}^T \in {^{\Psi}\Omega}.$$

*Then $^{\Psi}\Omega$ is said to be a generalized phase space or $\Psi$-space (PSI – space) [27-29].*

**Remark 1**

In case of the finite dimensional phase space, let us use notation $_N\vec{\xi} = \left\{\vec{r},\vec{v},\dot{\vec{v}},..., \overset{(N-2)}{\vec{v}}\right\}^T \in {^{\Psi}_N\Omega}$. For example, the classical phase space is noted as $^{\Psi}_2\Omega$ and has the following elements $_2\vec{\xi} = \{\vec{r},\vec{v}\}^T$.

***Definition 2.*** *Let a kinematic point be characterized at some initial time $t_0 \in T$ by the state $\vec{\xi}_0 \in {^{\Psi}\Omega}$, where*

$$\vec{\xi}_0 = \{\vec{r}_0, \vec{v}_0, \dot{\vec{v}}_0, \ddot{\vec{v}}_0, \ldots\}^T \in {}^{\Psi}\Omega.$$

*The generalized phase trajectory of the point in the space ${}^{\Psi}\Omega$ is defined as*

$$\vec{\xi}(t) = \begin{pmatrix} \vec{r}(t) \\ \vec{v}(t) \\ \dot{\vec{v}}(t) \\ \ldots \end{pmatrix} = \begin{pmatrix} \vec{r}_0 + \vec{v}_0(t-t_0) + \dfrac{\dot{\vec{v}}_0(t-t_0)^2}{2!} + \dfrac{\ddot{\vec{v}}_0(t-t_0)^3}{3!} + \ldots \\ \vec{v}_0 + \dot{\vec{v}}_0(t-t_0) + \dfrac{\ddot{\vec{v}}_0(t-t_0)^2}{2!} + \dfrac{\dddot{\vec{v}}_0(t-t_0)^3}{3!} + \ldots \\ \dot{\vec{v}}_0 + \ddot{\vec{v}}_0(t-t_0) + \dfrac{\dddot{\vec{v}}_0(t-t_0)^2}{2!} + \dfrac{\vec{v}_0^{(4)}(t-t_0)^3}{3!} + \ldots \\ \ldots \end{pmatrix}, \qquad (1.1)$$

*provided that series (1.1) converge.*

Note that, on a generalized phase trajectory, the vector quantities $\vec{r}_0, \vec{v}_0, \dot{\vec{v}}_0, \ddot{\vec{v}}_0, \ldots$ are dependent; i.e.,

$$\vec{v}_0 = \left.\dfrac{d\vec{r}}{dt}\right|_{t=t_0},\ \dot{\vec{v}}_0 = \left.\dfrac{d\vec{v}}{dt}\right|_{t=t_0} = \left.\dfrac{d^2\vec{r}}{dt^2}\right|_{t=t_0},\ldots \qquad (1.2)$$

Thus, if conditions (1.2) hold, then expression (1.1) contains Taylor series converging for sufficiently smooth trajectories.

***Corollary 1.***
 (i) *It follows from (1.1) and (1.2) that the series in (1.1) do not converge for all points of ${}^{\Psi}\Omega$. Consider a subspace ${}^{\Psi}\Omega'$ with elements $\vec{\xi} \in {}^{\Psi}\Omega' \subset {}^{\Psi}\Omega$ for which (1.1) are the Taylor series.*
 (ii) *Since the Taylor series expansion is unique, only one generalized phase trajectory (1.1) passes through each point of the subspace ${}^{\Psi}\Omega'$.*
 (iii) *Generalized phase trajectories in ${}^{\Psi}\Omega'$ do not intersect.*
 (iv) *Given a state $\vec{\xi} = \vec{\xi}(t) \in {}^{\Psi}\Omega'$, expression (1.1) puts it in correspondence with a state $\vec{\xi}_0 = \vec{\xi}(t_0) \in {}^{\Psi}\Omega'$; i.e., it is a mapping, which is referred to as the Taylor mapping. This mapping is one-to-one.*

***Definition 3.*** *Let $\vec{\xi}(t)$ is a generalized phase trajectory (1.1), with tangent vectors in each point of it*

$$\vec{u}_{\xi}(t) = \frac{d\vec{\xi}(t)}{dt} = \begin{pmatrix} \dfrac{d\vec{r}(t)}{dt} \\ \dfrac{d\vec{v}(t)}{dt} \\ \dfrac{d\dot{\vec{v}}(t)}{dt} \\ \ldots \end{pmatrix} = \begin{pmatrix} \vec{v}(t) \\ \dot{\vec{v}}(t) \\ \ddot{\vec{v}}(t) \\ \ldots \end{pmatrix} = \begin{pmatrix} \vec{v}_0 + \dot{\vec{v}}_0(t-t_0) + \ddot{\vec{v}}_0 \dfrac{(t-t_0)^2}{2} + \ldots \\ \dot{\vec{v}}_0 + \ddot{\vec{v}}_0(t-t_0) + \dddot{\vec{v}}_0 \dfrac{(t-t_0)^2}{2} + \ldots \\ \ddot{\vec{v}}_0 + \dddot{\vec{v}}_0(t-t_0) + \ddddot{\vec{v}}_0 \dfrac{(t-t_0)^2}{2} + \ldots \\ \ldots \end{pmatrix}, \quad (1.3)$$

*then we denote vector $\vec{u}_{\xi}(t)$ as generalized vector of the velocity of the physical system state change.*

As the single generalized phase trajectory passes through each point $\vec{\xi}$ of the space $^{\Psi}\Omega'$, then there is *the only way* to define the generalized vector of the velocity of the physical system state change in each point $\vec{\xi}$. Therefore, instead of the notation $\vec{u}_{\xi}(t)$, we use the notation $\vec{u}(\vec{\xi})$. As a result, the vector field of the generalized velocity $\vec{u} = \vec{u}(\vec{\xi}) = \{\vec{v}, \dot{\vec{v}}, \ddot{\vec{v}}, \ldots\}^T$ is plotted in the space $^{\Psi}\Omega'$.

***Theorem 1*** *(generalized Liouville's theorem). The Taylor mapping taking a domain $\omega_1 \subset {}^{\Psi}\Omega'$ to a domain $\omega_2 \subset {}^{\Psi}\Omega'$ has a Jacobian determinant equal to unity; i.e.,*

$$J = \left| \frac{\partial(\vec{r}^{(2)}, \vec{v}^{(2)}, \dot{\vec{v}}^{(2)}, \ddot{\vec{v}}^{(2)}, \ldots)}{\partial(\vec{r}^{(1)}, \vec{v}^{(1)}, \dot{\vec{v}}^{(1)}, \ddot{\vec{v}}^{(1)}, \ldots)} \right| = 1, \quad (1.4)$$

*where* $\vec{\xi}^{(2)} = \{\vec{r}^{(2)}, \vec{v}^{(2)}, \dot{\vec{v}}^{(2)}, \ddot{\vec{v}}^{(2)}, \ldots\}^T \in \omega_2$, $\vec{\xi}^{(1)} = \{\vec{r}^{(1)}, \vec{v}^{(1)}, \dot{\vec{v}}^{(1)}, \ddot{\vec{v}}^{(1)}, \ldots\}^T \in \omega_1$, $\vec{r}^{(2)} = \vec{r}^{(2)}(\vec{r}^{(1)}, \vec{v}^{(1)}, \dot{\vec{v}}^{(1)}, \ddot{\vec{v}}^{(1)}, \ldots), \vec{v}^{(2)} = \vec{v}^{(2)}(\vec{v}^{(1)}, \dot{\vec{v}}^{(1)}, \ddot{\vec{v}}^{(1)}, \ldots), \ldots$

***Corollary 2****. It is true that:*

$$J_1 = \left| \frac{\partial(\vec{v}^{(2)}, \dot{\vec{v}}^{(2)}, \ddot{\vec{v}}^{(2)}, \ldots)}{\partial(\vec{v}^{(1)}, \dot{\vec{v}}^{(1)}, \ddot{\vec{v}}^{(1)}, \ldots)} \right| = 1, \; J_2 = \left| \frac{\partial(\dot{\vec{v}}^{(2)}, \ddot{\vec{v}}^{(2)}, \ldots)}{\partial(\dot{\vec{v}}^{(1)}, \ddot{\vec{v}}^{(1)}, \ldots)} \right| = 1, \ldots \quad (1.5)$$

**Remark 2**

In the classical phase space $^{\Psi}_2\Omega$, the trajectory by which the phase space point $_2\vec{\xi}_0 = \{\vec{r}_0, \vec{v}_0\}^T$ transforms into the point $_2\vec{\xi}_1 = \{\vec{r}_1, \vec{v}_1\}^T$ satisfies Cauchy problem for the Hamiltonian equations with the initial values $_2\vec{\xi}_0 = \{\vec{r}_0, \vec{v}_0\}^T$.

In the generalized phase space $^{\Psi}\Omega'$ the point is defined by an infinite axis set $\vec{\xi}_0 = \{\vec{r}_0, \vec{v}_0, \dot{\vec{v}}_0, \ddot{\vec{v}}_0, \ldots\}^T$, by which the trajectory is plotted uniquely in view of (1.1) with no need to solve Cauchy problem for the Hamiltonian equations. Thus, the trajectory equations in the generalized phase space are analytical functions mathematically.

The Cauchy problem for the trajectory $\vec{\xi}(t)$ is of the form (1.1), (1.3):

$$\begin{cases} \dfrac{d\vec{\xi}(t)}{dt} = \vec{u}\left(\vec{\xi}(t)\right), \\ \vec{\xi}(t_0) = \vec{\xi}_0. \end{cases} \qquad (1.3A)$$

**Remark 3**

An interesting question is the form and maximum allowed size of the boundary $\Gamma'$ of the region $\omega$ in the space $^{\Psi}\Omega'$. Such a boundary imposes limitations on the allowed values of the parameters $\vec{r}, \vec{v}, \dot{\vec{v}}, \ldots$ of a physical system.

Mathematically, such limitations are determined by the convergence of Taylor's series, standing in the definition of the generalized phase (1.1). Thus, it is possible to estimate the radius of the convergence of the series for each point of the generalized phase space and so to define the boundary $\Gamma'$.

From the physical standpoint, it is known from relativity theory that the limitation on the velocity modulus is $|\vec{v}| \leq c$, where $c$ is the speed of light.

The question of the acceleration finiteness $|\dot{\vec{v}}|$ may be interpreted as a question of the maximum possible force in the Universe or the maximum synchrotron emission power, which is proportional to $|\dot{\vec{v}}|^2$ as it is known [24]. The finiteness of the rest motion characteristics $\ddot{v}, \dddot{v}, \ldots$ and the question of their physical interpretation remains open for now.

In many physics applications, such as accelerator physics, not the phase domain itself is considered but its projections on axis planes. For example, the projection planes $(x, v_x)$, $(y, v_y)$, $(z, v_z)$ are considered for the 6D phase space $^{\Psi}_2\Omega'$ with the elements $_2\vec{\xi} = \{x, y, z, v_x, v_y, v_z\}^T$. In the mentioned projection planes the phase «area» conservation is considered or, in terms of accelerator physics, beam emittance. Therefore, let us consider the issue of conservation of projection volumes in the generalized phase space.

***Definition 4.*** *The X-, Y-, and Z-projection phase spaces are the respective spaces $\omega_x$, $\omega_y$, $\omega_z$ with elements $\eta^{(x)} = \{x, v^{(x)}, \dot{v}^{(x)}, \ddot{v}^{(x)}, \ldots\}^T \in \omega_x$, $\eta^{(y)} = \{y, v^{(y)}, \dot{v}^{(y)}, \ddot{v}^{(y)}, \ldots\}^T \in \omega_y$, $\eta^{(z)} = \{z, v^{(z)}, \dot{v}^{(z)}, \ddot{v}^{(z)}, \ldots\}^T \in \omega_z$, where $\vec{\xi} = \{x, y, z, v^{(x)}, v^{(y)}, v^{(z)}, \dot{v}^{(x)}, \ldots\}^T \in {}^{\Psi}\Omega'$.*

***Definition 5.*** *A mapping of the form*

$$\eta^{\left(x^{(i)}\right)}(t) = \begin{pmatrix} x^{(i)}(t) \\ v^{\left(x^{(i)}\right)}(t) \\ \dot{v}^{\left(x^{(i)}\right)}(t) \\ \ldots \end{pmatrix} = \begin{pmatrix} x_0^{(i)} + v_0^{\left(x^{(i)}\right)}(t-t_0) + \dfrac{\dot{v}_0^{\left(x^{(i)}\right)}(t-t_0)^2}{2!} + \dfrac{\ddot{v}_0^{\left(x^{(i)}\right)}(t-t_0)^3}{3!} + \ldots \\ v_0^{\left(x^{(i)}\right)} + \dot{v}_0^{\left(x^{(i)}\right)}(t-t_0) + \dfrac{\ddot{v}_0^{\left(x^{(i)}\right)}(t-t_0)^2}{2!} + \dfrac{\dddot{v}_0^{\left(x^{(i)}\right)}(t-t_0)^3}{3!} + \ldots \\ \dot{v}_0^{\left(x^{(i)}\right)} + \ddot{v}_0^{\left(x^{(i)}\right)}(t-t_0) + \dfrac{\dddot{v}_0^{\left(x^{(i)}\right)}(t-t_0)^2}{2!} + \dfrac{\ddddot{v}_0^{\left(x^{(i)}\right)}(t-t_0)^3}{3!} + \ldots \\ \ldots \end{pmatrix} \qquad (1.6)$$

is called the *X-, Y-,* and *Z-projection Taylor mappings* for $i = 1, 2, 3$, respectively.

**Theorem 2.** *The X-, Y-, and Z-projection Taylor mappings have Jacobian determinants equal to unity; i.e.,*

$$\left|J^{(x^{(i)})}\right| = \left|\frac{\partial\left(x_2^{(i)}, v_2^{(x^{(i)})}, \dot{v}_2^{(x^{(i)})}, \ddot{v}_2^{(x^{(i)})}, \ldots\right)}{\partial\left(x_1^{(i)}, v_1^{(x^{(i)})}, \dot{v}_1^{(x^{(i)})}, \ddot{v}_1^{(x^{(i)})}, \ldots\right)}\right| = 1, \quad x^{(1)} = x, \ x^{(2)} = y, \ x^{(3)} = z \qquad (1.7)$$

*where* $\eta_2^{(x^{(i)})} = \left\{x_2^{(i)}, v_2^{(x^{(i)})}, \dot{v}_2^{(x^{(i)})}, \ddot{v}_2^{(x^{(i)})}, \ldots\right\}^T$, $\eta_1^{(x^{(i)})} = \left\{x_1^{(i)}, v_1^{(x^{(i)})}, \dot{v}_1^{(x^{(i)})}, \ddot{v}_1^{(x^{(i)})}, \ldots\right\}^T$, ...,

$x_2^{(i)} = x_2^{(i)}\left(x_1^{(i)}, v_1^{(x^{(i)})}, \dot{v}_1^{(x^{(i)})}, \ddot{v}_1^{(x^{(i)})}, \ldots\right)$, $v_2^{(x^{(i)})} = v_2^{(x^{(i)})}\left(v_1^{(x^{(i)})}, \dot{v}_1^{(x^{(i)})}, \ddot{v}_1^{(x^{(i)})}, \ldots\right)$, .... .

**Corollary 3.** *It is true that:*

$$\left|J_1^{(x^{(i)})}\right| = \left|\frac{\partial\left(v_2^{(x^{(i)})}, \dot{v}_2^{(x^{(i)})}, \ddot{v}_2^{(x^{(i)})}, \ldots\right)}{\partial\left(v_1^{(x^{(i)})}, \dot{v}_1^{(x^{(i)})}, \ddot{v}_1^{(x^{(i)})}, \ldots\right)}\right| = 1, \ \left|J_2^{(x^{(i)})}\right| = \left|\frac{\partial\left(\dot{v}_2^{(x^{(i)})}, \ddot{v}_2^{(x^{(i)})}, \ldots\right)}{\partial\left(\dot{v}_1^{(x^{(i)})}, \ddot{v}_1^{(x^{(i)})}, \ldots\right)}\right| = 1, \ldots \qquad (1.8)$$

**Theorem 3.** *If elementary transformations of the form*

$$\vec{\xi}(t) = \left\{\eta^{(x)}(t), \eta^{(y)}(t), \eta^{(z)}(t)\right\}^T =$$
$$= \left\{x(t), v^{(x)}(t), \dot{v}^{(x)}(t), \ldots, y(t), v^{(y)}(t), \dot{v}^{(y)}(t), \ldots, z(t), v^{(z)}(t), \dot{v}^{(z)}(t), \ldots\right\}^T$$

*are applied to vectors in the subspace* $^\Psi\Omega'$, *then the determinant of the Jacobian matrix (1.4) becomes*

$$|J| = \begin{vmatrix} J^{(x)} & 0 & 0 \\ 0 & J^{(y)} & 0 \\ 0 & 0 & J^{(z)} \end{vmatrix} = 1, \ i.e., \ J = J^{(x)} \oplus J^{(y)} \oplus J^{(z)}. \qquad (1.9)$$

In accelerator physics, plasma physics, at considering beams of charged particles, the velocity $\vec{v}$, acceleration $\dot{\vec{v}}$ and the rest high order accelerations $\ddot{\vec{v}}, \dddot{\vec{v}}, \ldots$ are put uniquely corresponding to each particle with the coordinate $\vec{r}$. Let us introduce the generalized phase space $^\Psi\Omega''$ for such systems.

**Definition 6.** *Let* $^\Psi\Omega''$ *denote the subspace of* $^\Psi\Omega'$ *in which*

$$\forall \vec{\xi}_1, \vec{\xi}_2 \in {^\Psi\Omega''} \ (\vec{r}_1 = \vec{r}_2 = \vec{r}) : \vec{v}_1 = \vec{v}_2, \dot{\vec{v}}_1 = \dot{\vec{v}}_2, \ddot{\vec{v}}_1 = \ddot{\vec{v}}_2, \ldots,$$

*where* $\vec{\xi}_1 = \{\vec{r}_1, \vec{v}_1, \dot{\vec{v}}_1, ...\}^T$, $\vec{\xi}_2 = \{\vec{r}_2, \vec{v}_2, \dot{\vec{v}}_2, ...\}^T$.

***Theorem 4.*** *The maximum dimension of the subspace* $^\Psi\Omega''$ *is 3.*

***Corollary 4.*** The dimensionality $n$ of the projection phase subspaces $\omega_x$, $\omega_y$, $\omega_z$ satisfies $n \geq 1$.

**Remark 4**

In accelerator physics, a particle beam is characterized usually by a 6D phase ellipse:

$$\frac{x^2}{\delta_x^2} + \frac{y^2}{\delta_y^2} + \frac{z^2}{\delta_z^2} + \frac{v_x^2}{\delta v_x^2} + \frac{v_y^2}{\delta v_y^2} + \frac{v_z^2}{\delta v_z^2} = 1 \qquad (1.10)$$

Emittance is a numerical characteristic of an accelerated beam of charged particles which is equal to the volume of the phase space (6D in general case) taken by the beam. Emittance is an important parameter of the beams obtained at particle accelerators as it determines largely the efficiency of further beam usage.

Points of the space $^\Psi\Omega''$ correspond to the beam particles (see Definition 5). From Theorem 4 and expression (1.10), it follows that points corresponding to the beam cannot fill the 6D volume of the ellipse (1.10), above that they do not fill the ellipse surface (1.10) as in this case, there is an infinite aggregate of velocity component variations of the following form:

$$\frac{v_x^2}{\delta v_x^2} + \frac{v_y^2}{\delta v_y^2} + \frac{v_z^2}{\delta v_z^2} = C,$$

corresponding to the same particle with the coordinate $x, y, z$, where

$$C = 1 - \frac{x^2}{\delta_x^2} - \frac{y^2}{\delta_y^2} - \frac{z^2}{\delta_z^2} = const.$$

**§ 2 Multi-particle system**

Let us demonstrate the qualitative distinction between the generalized phase space $^\Psi\Omega'$ and the classical phase space $^\Psi_2\Omega'$. For display purposes, let us consider a one-dimensional system of $N$ - particles. For simplicity, each $i$-th particle is described with the use of three components only in the generalized phase space $^\Psi_3\Omega'$: $x_i, v_i^{(x)}, \dot{v}_i^{(x)}$, $i = 1...N$. The other components $\ddot{v}_i^{(x)} = \dddot{v}_i^{(x)} = ... = 0$.

According to Theorem 2 and Definition 3, it is enough to consider only the $X$ - projection space $\omega_x$ with the $X$ - projection Taylor mapping (4).

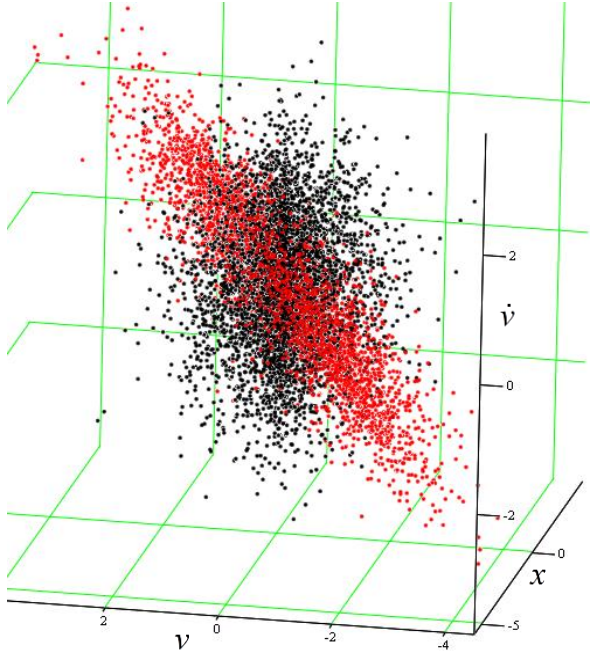

Fig. 1. Distribution of particles in the phase space $\omega_x$

Let us generate in random manner the distribution of $N$ - particles in the space $\omega_x$ (Fig. 1, black points).

Let this distribution (see Fig.1, black points) $\vec{\xi}_i$ corresponds to some initial moment of time $t = t_0 = 0$. According to transformation (1.6), in the moment of time $t = \tau$ the distribution is $\vec{\tilde{\xi}}_i$ (see Fig.1, red points) (2.1):

$$\vec{\xi}_i = \begin{pmatrix} x_i \\ v_i \\ \dot{v}_i \end{pmatrix}, \vec{\tilde{\xi}}_i = \begin{pmatrix} \tilde{x}_i \\ \tilde{v}_i \\ \dot{\tilde{v}}_i \end{pmatrix} = \begin{pmatrix} x_i + v_i \tau + \dfrac{\dot{v}_i \tau^2}{2!} \\ v_i + \dot{v}_i \tau \\ \dot{v}_i \end{pmatrix}, (2.1)$$

where $i = 1, ..., N$. According to Theorem 2, the Jacobian matrix determinant of such mapping (2.1) equals to one, it is true:

$$\left| J^{(x)} \right| = \left| \frac{\partial (\tilde{x}, \tilde{v}, \dot{\tilde{v}})}{\partial (x, v, \dot{v})} \right| = \begin{vmatrix} 1 & \tau & \dfrac{\tau^2}{2} \\ 0 & 1 & \tau \\ 0 & 0 & 1 \end{vmatrix} = 1 \qquad (2.2)$$

As well, according to Corollary 3, the determinants $J_1^{(x)}$, $J_2^{(x)}$ equal to one too, it is true (1.8):

$$\left| J_1^{(x)} \right| = \left| \frac{\partial (\tilde{v}, \dot{\tilde{v}})}{\partial (v, \dot{v})} \right| = \begin{vmatrix} 1 & \tau \\ 0 & 1 \end{vmatrix} = 1, \quad \left| J_2^{(x)} \right| = \left| \frac{\partial (\dot{\tilde{v}})}{\partial (\dot{v})} \right| = 1. \qquad (2.3)$$

Let us check on statements (2.2) and (2.3) directly through the statistical analysis of the particle distributions $\vec{\xi}_i$ and $\vec{\tilde{\xi}}_i$. In Figs.2, 3 the projections $(x, v)$, $(v, \dot{v})$ and $(x, \dot{v})$ are shown of the distributions $\vec{\xi}_i$ and $\vec{\tilde{\xi}}_i$ respectively.

In order to evaluate the projection surface area in Figs. 2, 3 let us use the RMS-emittance method, which is used frequently in accelerator physics [25]. In Figs. 2,3 RMS beam ellipses are blue. The ellipse equation for the projection $(x, v)$ is as follows

$$\gamma x^2 + 2\alpha xv + \beta v^2 = \varepsilon, \qquad (2.4)$$

where $\alpha, \beta, \gamma$ are Courant-Snyder parameters, and $\varepsilon$ − emittance, which is related to the ellipse surface area $s$ as follows $s = \pi \varepsilon$. For the projections $(x, v)$ and $(x, \dot{v})$, the equations have analogue form.

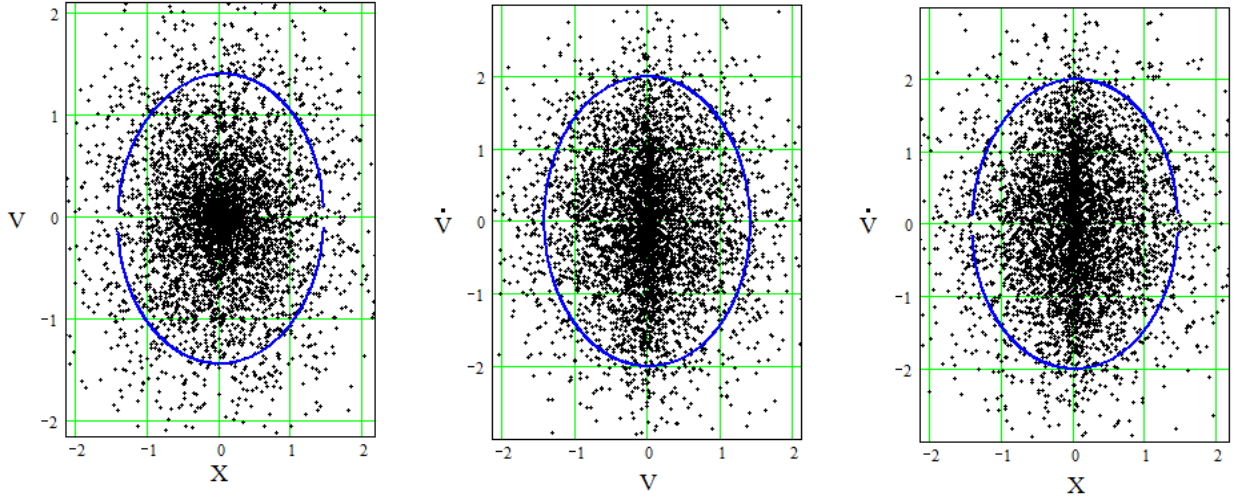

Fig. 2. Projections of the distribution $\vec{\xi}_i$ on the phase planes $(x,v)$, $(v,\dot{v})$ and $(x,\dot{v})$

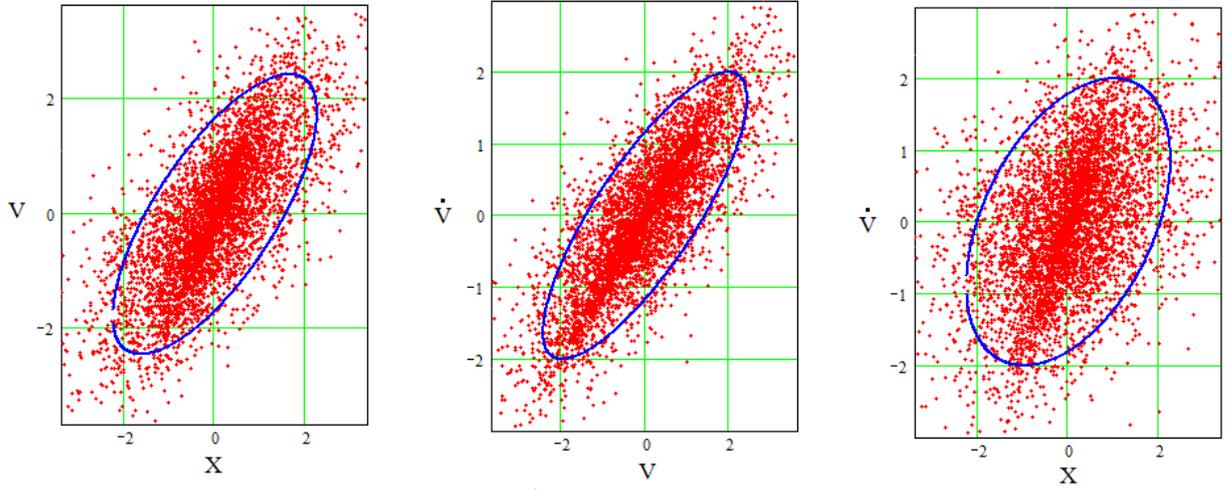

Fig. 3. Projections of the distribution $\vec{\tilde{\xi}}_i$ on the phase planes $(x,v)$, $(x,v)$ and $(v,\dot{v})$

In Table 1, the values for the surface areas are listed for the projections $(x,v)$, $(x,v)$, $(v,\dot{v})$ as well as standard deviations of the accelerations $\sigma_{\dot{v}}$ for the distributions $\vec{\xi}_i$ and $\vec{\tilde{\xi}}_i$.

Table 1

| $\vec{\xi}$ \ $s$ | $(x,v)$ | $(x,v)$ | $(v,\dot{v})$ | $2\sigma_{\dot{v}}$ |
|---|---|---|---|---|
| $\vec{\xi}_i$ | 6.436 | 8.882 | 9.024 | 1.991 |
| $\vec{\tilde{\xi}}_i$ | 11.922 | 8.882 | 12.755 | 1.991 |

As it is seen from Table 1, the surface are of the projection $(x,v)$ is conserved, which corresponds to the first congruence in expression (2.3), i.e. $\left|J_1^{(x)}\right|=1$. The conservation (Table 1) of the projection on the acceleration axis $2\sigma_{\dot{v}}$ corresponds to the second congruence of

expression (2.3), i.e. $\left|J_2^{(x)}\right|=1$. From Table 1 it is seen, that the projections $(x,v)$ and $(v,\dot{v})$ are not conserved, which is also in agreement with Theorem 2.

Let us check on condition (2.2) on the conservation of the full phase volume in the three-dimensional space $(x,v,\dot{v})$. In order to check condition (2.2), it is needed to calculate the volume of two ellipses (see Fig.1) corresponding to the distributions $\vec{\xi}_i$ and $\vec{\tilde{\xi}}_i$. If the principal semiaxes of the ellipse $a,b,c$ are found, then the volume of the ellipse may be found with the formula $V=\frac{4}{3}\pi abc$. To find $a,b,c$, let us use the principal component analysis (PCA) [26]. The covariance matrix for the distributions $\vec{\xi}_i$ and $\vec{\tilde{\xi}}_i$ are of the form:

$$Cov(\vec{\xi})=\begin{pmatrix} \operatorname{cov}(x,x) & \operatorname{cov}(x,v) & \operatorname{cov}(x,\dot{v}) \\ \operatorname{cov}(v,x) & \operatorname{cov}(v,v) & \operatorname{cov}(v,\dot{v}) \\ \operatorname{cov}(\dot{v},x) & \operatorname{cov}(\dot{v},v) & \operatorname{cov}(\dot{v},\dot{v}) \end{pmatrix} \simeq \begin{pmatrix} 0.52 & 0.008 & 0.007 \\ 0.008 & 0.504 & -0.006 \\ 0.007 & -0.006 & 0.991 \end{pmatrix}, \quad (2.5)$$

$$Cov(\vec{\tilde{\xi}}) \simeq \begin{pmatrix} 1.288 & 1.005 & 0.496 \\ 1.005 & 1.482 & 0.985 \\ 0.496 & 0.985 & 0.991 \end{pmatrix}.$$

Transiting into basis out of eigenvectors of the matrix (2.5) becomes diagonal. The eigenvalues of the matrixes are on its diagonals, and they are expressed via the standard deviations $\sigma$, i.e.

$$Cov(\vec{\xi}')=\begin{pmatrix} \sigma_{x'}^2 & 0 & 0 \\ 0 & \sigma_{v'}^2 & 0 \\ 0 & 0 & \sigma_{\dot{v}'}^2 \end{pmatrix} \simeq \begin{pmatrix} 0.52 & 0 & 0 \\ 0 & 0.501 & 0 \\ 0 & 0 & 0.991 \end{pmatrix}, \quad (2.6)$$

$$Cov(\vec{\tilde{\xi}}') \simeq \begin{pmatrix} 0.133 & 0 & 0 \\ 0 & 0.654 & 0 \\ 0 & 0 & 2.974 \end{pmatrix}.$$

As $a=2\sigma_{x'}$, $b=2\sigma_{v'}$, $c=2\sigma_{\dot{v}'}$, then, from (2.6) it follows that the volume of the ellipse corresponding to the distribution $\vec{\xi}_i$ is $V \simeq 17.076$. By analogue, the volume of the ellipse corresponding to the distribution $\vec{\tilde{\xi}}_i$ is $V' \simeq 17.076$. That is $V \simeq V'$, which is in agreement with statement (2.2) $\left|J^{(x)}\right|=1$.

**Remark 5**

Note, that motion trajectories (2.1), along which the system of $N$ particles moves, satisfies the Hamiltonian equations. According to classical Liouville's theorem for transformations (2.1), the phase volume in the projection $(x,v)$ must be conserved. Nevertheless, from the considered example, it follows that in the projection $(x,v)$ the phase volume is not conserved. According to Figs. 2, 3 and Table 1, in the projection $(x,v)$ the phase volume almost doubles from $S_{(x,v)}(\vec{\xi})=6.436$ to $S_{(x,v)}(\vec{\tilde{\xi}})=11.922$.

The said non-conservation of the phase volume is caused by the fact that the considered system of $N$ - particles is not a Hamiltonian system. In the Hamiltonian system, the set of kinematic characteristics $\vec{v}, \dot{\vec{v}}, \ddot{\vec{v}},...$ uniquely corresponds to each particle with a coordinate $\vec{r}$, i.e. the space, where the Hamiltonian system is defined, is the space $^{\Psi}\Omega''$ (see Definition 5). It is the space $^{\Psi}\Omega''$ where the phase volume $(x,v)$ is conserved.

In the space $^{\Psi}_{3}\Omega'$, two particles $A$ and $B$ may exist with the same coordinates $\vec{r}_A = \vec{r}_B$ and velocities $\vec{v}_A = \vec{v}_B$, but with different accelerations $\dot{\vec{v}}_A \neq \dot{\vec{v}}_B$. Mathematically, different points correspond to the particles $A$ and $B$ in the generalized phase space $^{\Psi}_{3}\Omega'$, as $\dot{\vec{v}}_A \neq \dot{\vec{v}}_B$. In the classical phase space $^{\Psi}_{2}\Omega'$, points $A$ and $B$ coincide, as $\vec{r}_A = \vec{r}_B$ and $\vec{v}_A = \vec{v}_B$. At moving along trajectories (2.1), the points $A$ and $B$ in the space $^{\Psi}_{2}\Omega'$ diverge, as $\dot{\vec{v}}_A \neq \dot{\vec{v}}_B$, which leads to the increase in the classical phase volume $(\vec{r},\vec{v})$. Analogue reasoning is true for reverse motion, which leads to the classical phase volume $(\vec{r},\vec{v})$ compresses. The generalized phase volume $(\vec{r},\vec{v},\dot{\vec{v}})$ stays unchanged in both cases.

If the space $^{\Psi}\Omega''$ corresponds to a physical system, then no «compression» or «extension» of the classical phase volume occurs and classical Liouville's theorem on the phase volume $(\vec{r},\vec{v})$ conservation is fulfilled. The space $^{\Psi}\Omega'$ corresponds to a dissipative system in general case, and in this space the classical phase volume $(\vec{r},\vec{v})$ may be conserved or not, but the generalized phase volume $(\vec{r},\vec{v},\dot{\vec{v}},\ddot{\vec{v}},.....)$ is always conserved.

## §3 Liouville's generalized equation

Let us define the generalized phase space-time [27-29].

**Definition 7.** *Let us call the space* $^{\Psi}\Omega_T = {}^{\Psi}\Omega \times T$, *where* $T = \{t \in \mathbb{R} : -\infty < t < +\infty\}$ *the generalized phase space-time.*

In the generalized phase space-time $^{\Psi}\Omega_T$, let the probability density functions $f_k(\vec{r},\vec{v},\dot{\vec{v}},\ddot{\vec{v}},...,t)$ be defined, $k = 1, 2,...$

$$\begin{aligned} &f_1(\vec{r},t), \\ &f_2(\vec{r},\vec{v},t), \\ &f_3(\vec{r},\vec{v},\dot{\vec{v}},t), \\ &... \end{aligned} \quad (3.1)$$

where $\vec{r},\vec{v},\dot{\vec{v}},\ddot{\vec{v}},...$ are independent variables.

Probability density functions (3.1) satisfy the relations [20, 21, 27-29]

$$f_0(t) \overset{\text{det}}{=} N(t) = \int\limits_{(\infty)} f_1(\vec{r},t) d^3r = \int\limits_{(\infty)}\int\limits_{(\infty)} f_2(\vec{r},\vec{v},t) d^3r d^3v = ... =$$
$$= ... = \int\limits_{(\infty)}\int\limits_{(\infty)}\int\limits_{(\infty)} ... f_\infty(\vec{r},\vec{v},\dot{\vec{v}},...,t) d^3r d^3v d^3\dot{v}... \qquad (3.2)$$

where $N(t)$ is normalization factor or a number of particles, which can be non-integer in general case [20,21]. Assume the functions $f_k({}_k\vec{\xi},t)$ satisfy condition [20, 21]

$$\lim_{\xi\to\infty}\left[{}_k\xi f_k({}_k\vec{\xi},t)\right] = 0. \qquad (3.3)$$

In the middle of the 20th century, Vlasov wrote the generalized continuity equation for the probability density functions $f_\infty(\vec{\xi},t)$ [20, 21, 27-29]

$$\frac{\partial f_\infty(\vec{\xi},t)}{\partial t} + \text{div}_\xi\left[f_\infty(\vec{\xi},t)\vec{u}(\vec{\xi})\right] = 0, \qquad (3.4)$$

where $\vec{u}(\vec{\xi})$ – the generalized velocity vector, and $\text{div}_\xi \overset{\text{det}}{=} \text{div}_r + \text{div}_v + \text{div}_{\dot{v}} + ...$. Formally applying the integration over the domains $\omega_v, \omega_{\dot{v}}, \omega_{\ddot{v}},...$ to equation (3.4) as well as the divergence theorem and taking into account condition (3.3), Vlasov obtained an infinite chain of equations for the functions $f_k({}_k\vec{\xi},t)$

$$\frac{\partial f_1(\vec{r},t)}{\partial t} + \text{div}_r \int\limits_{(\infty)} f_2(\vec{r},\vec{v},t)\vec{v} d^3v = 0,$$
$$\frac{\partial f_2(\vec{r},\vec{v},t)}{\partial t} + \text{div}_r\left[f_2(\vec{r},\vec{v},t)\vec{v}\right] + \text{div}_v \int\limits_{(\infty)} f_3(\vec{r},\vec{v},\dot{\vec{v}},t)\dot{\vec{v}} d^3\dot{v} = 0, \qquad (3.5)$$
$$...$$
$$\frac{\partial f_n\left(\vec{r},\vec{v},...,\overset{(n-1)}{\vec{r}},t\right)}{\partial t} + \text{div}_r\left[f_n\left(\vec{r},\vec{v},...,\overset{(n-1)}{\vec{r}},t\right)\vec{v}\right] + ... + \text{div}_{\overset{(n-1)}{r}} \int\limits_{(\infty)} f_{n+1}\left(\vec{r},\vec{v},...,\overset{(n)}{\vec{r}},t\right)\overset{(n)}{\vec{r}} d^3 \overset{(n)}{\vec{r}} = 0,$$
$$...$$

The second equation in chain (3.5) is known as the Vlasov equation for the self-consistent field.

Rewrite equation (3.4) in the form

$$\frac{\partial f_\infty}{\partial t} + f_\infty \text{div}_\xi[\vec{u}] + \vec{u}\nabla_\xi f_\infty = 0 \qquad (3.6)$$

The vector $\vec{u}(\vec{\xi})$ is, according to definition (1.3), of the form:

$$\vec{u}(\vec{\xi}) = \begin{pmatrix} \vec{v} \\ \dot{\vec{v}} \\ \ddot{\vec{v}} \\ \cdots \end{pmatrix}, \quad \vec{\xi} = \begin{pmatrix} \vec{r} \\ \vec{v} \\ \dot{\vec{v}} \\ \cdots \end{pmatrix},$$

consequently,

$$\operatorname{div}_{\xi}\left[\vec{u}(\vec{\xi})\right] = \operatorname{div}_{r}[\vec{v}] + \operatorname{div}_{v}[\dot{\vec{v}}] + \operatorname{div}_{\dot{v}}[\ddot{\vec{v}}] + \ldots = 0. \qquad (3.7)$$

Considering (3.7), equation (3.6) is of the form:

$$\frac{\partial f_{\infty}}{\partial t} + \vec{u}\nabla_{\xi} f_{\infty} = 0, \qquad (3.8)$$

or

$$\frac{\partial f_{\infty}}{\partial t} + \vec{v}\nabla_{r} f_{\infty} + \dot{\vec{v}}\nabla_{v} f_{\infty} + \ddot{\vec{v}}\nabla_{\dot{v}} f_{\infty} + \ldots = 0.$$

Equation (3.8) has solutions in the form of the characteristics $\vec{\xi}(t)$, while the characteristics are generalized phase trajectories (1.1), satisfying the Cauchy problem (1.3A):

$$\begin{cases} \dfrac{d\vec{\xi}(t)}{dt} = \vec{u}(\vec{\xi}(t)) \\ \vec{\xi}(t_0) = \vec{\xi}_0 \end{cases} \qquad (3.9)$$

At that, according to (3.8), the following is fulfilled:

$$f_{\infty}(\vec{\xi}(t), t) = Const \qquad (3.10)$$

***Definition 8.*** *Let us call equation (3.8)/(3.4) generalized Liouville's equation in the generalized phase space.*

From the above said, it follows that the Vlasov equation chain (3.5) is an equivalent expression of the generalized Liouville's equation (3.8)/(3.4) at transition to the finite dimensional phase spaces ${}^{\Psi}_{n}\Omega'$.

In paper [6], it is shown that the entropy $H_n$ equation chain is associated to the Vlasov infinite equation chain (3.5):

$$\frac{dH_1}{dt} = \langle Q_1 \rangle,$$
$$\frac{dH_2}{dt} = \langle\langle Q_2 \rangle\rangle,$$
$$...$$
$$\frac{dH_n}{dt} = \langle...\langle Q_n \rangle...\rangle,$$
$$...$$

(3.11)

where

$$\Pi_n \stackrel{det}{=} \frac{d_n}{dt} = \frac{\partial}{\partial t} + (\dot{\vec{r}}, \nabla_r) + ... + \left(\stackrel{(n-1)}{\vec{r}}, \nabla_{\stackrel{(n-2)}{r}}\right) + \left(\left\langle\stackrel{(n)}{\vec{r}}\right\rangle, \nabla_{\stackrel{(n-1)}{r}}\right) = \frac{\partial}{\partial t} + (\vec{u}_n, \nabla_{\xi_n}),$$

$$S_n(\vec{r}, \vec{v}, ..., t) \stackrel{det}{=} -\ln f_n(\vec{r}, \vec{v}, ..., t),$$

$$H_1(t) \stackrel{det}{=} -\frac{1}{N} \int_{(\infty)} f_1(\vec{r}, t) \ln f_1(\vec{r}, t) d^3r = \frac{1}{N} \int_{(\infty)} f_1(\vec{r}, t) S_1(\vec{r}, t) d^3r = \langle S_1 \rangle(t),$$

$$H_2(t) \stackrel{det}{=} -\frac{1}{N} \int_{(\infty)}\int_{(\infty)} f_2(\vec{r}, \vec{v}, t) \ln f_2(\vec{r}, \vec{v}, t) d^3r d^3v = \langle\langle S_2 \rangle\rangle(t),$$

(3.12)

$$...$$

$$H_n(t) \stackrel{det}{=} \langle...\langle S_n \rangle...\rangle(t),$$

$$...$$

$$Q_1 \stackrel{det}{=} \operatorname{div}_r \langle \vec{v} \rangle, \quad Q_2 \stackrel{det}{=} \operatorname{div}_v \langle \dot{\vec{v}} \rangle, \quad ..., \quad Q_n \stackrel{det}{=} \operatorname{div}_{\stackrel{(n-1)}{r}} \left\langle\stackrel{(n)}{\vec{r}}\right\rangle,$$

where average values are defined as

$$N(t)\langle \vec{r} \rangle(t) \stackrel{det}{=} \int_{(\infty)} f_1(\vec{r}, t) \vec{r} d^3r,$$

$$f_1(\vec{r}, t)\langle \vec{v} \rangle(\vec{r}, t) \stackrel{det}{=} \int_{(\infty)} f_2(\vec{r}, \vec{v}, t) \vec{v} d^3v,$$

$$N(t)\langle\langle \vec{v} \rangle\rangle(t) \stackrel{det}{=} \int_{(\infty)}\int_{(\infty)} f_2(\vec{r}, \vec{v}, t) \vec{v} d^3r d^3v,,$$

(3.13)

$$f_2(\vec{r}, \vec{v}, t)\langle \dot{\vec{v}} \rangle(\vec{r}, \vec{v}, t) \stackrel{det}{=} \int_{(\infty)} f_3(\vec{r}, \vec{v}, \dot{\vec{v}}, t) \dot{\vec{v}} d^3\dot{v},$$

$$f_1(r, t)\langle\langle \dot{\vec{v}} \rangle\rangle(\vec{r}, t) \stackrel{det}{=} \int_{(\infty)}\int_{(\infty)} f_3(\vec{r}, \vec{v}, \dot{\vec{v}}, t) \dot{\vec{v}} d^3v d^3\dot{v},$$

$$N(t)\langle\langle\langle \dot{\vec{v}} \rangle\rangle\rangle(t) \stackrel{det}{=} \int_{(\infty)}\int_{(\infty)}\int_{(\infty)} f_3(\vec{r}, \vec{v}, \dot{\vec{v}}, t) \dot{\vec{v}} d^3r d^3v d^3\dot{v},$$

$$...$$

The values $Q_n$ determine the sources of the entropy $H_n$ changes. Generalized Liouville's equation (3.8) may formally be rewritten via the generalized entropy $H_\infty$ of the infinite dimensional phase space $^\Psi\Omega'$

$$H_\infty \stackrel{det}{=} \langle...\langle S_\infty\rangle...\rangle,\ S_\infty \stackrel{det}{=} -\ln f_\infty,\ Q_\infty \stackrel{det}{=} \text{div}_\xi \vec{u}.$$

***Theorem 5.*** *Let function $f_\infty$ satisfy the equation (3.8), then the function $H_\infty$ satisfy the equation:*

$$\frac{d}{dt}\left[N(t)H_\infty(t)\right] = N(t)\langle...\langle Q_\infty\rangle...\rangle(t),$$

***Corollary 5.*** *If the number of particles is constant, that is $N(t) = const$, then theorem 5 results in the function of entropy $H_\infty$ satisfying the equation*

$$\frac{dH_\infty}{dt} = \langle...\langle Q_\infty\rangle...\rangle.$$

From (3.7), it follows that

$$Q_\infty = \text{div}_\xi \vec{u} = 0. \tag{3.14}$$

Taking into account (3.14) and Corollary 5, we obtain for the constant number of particles the following

$$\frac{dH_\infty}{dt} = 0. \tag{3.15}$$

From (3.15), it follows that the generalized entropy is a constant value $H_\infty = const$, as the generalized sources $Q_\infty$ equal to zero according to (3.14). At transition to the finite dimensional generalized phase spaces $^\Psi_n\Omega'$, the entropies $H_n$ get changed (3.11) as the sources $Q_n$ (3.12) are distinct from zero in general case.

In the infinite dimensional generalized phase space $^\Psi\Omega'$, there are no sources changing the entropy $Q_\infty = 0$, as the generalized phase trajectories $\vec{\xi}(t)$ are disjoint and $\text{div}_\xi \vec{u} = 0$. The transition to a lower dimensionality of the phase space happens as a result of averaging over kinematic characteristics (3.13).

The result of the transition to a lower dimensionality of the phase space is the concurrence of the «averaged» phase trajectories, as «projecting» happens of the infinite dimensional object into the finite dimensional space. The concurrence of the averaged phase trajectories leads to the existence of the sources $Q_n$ distinct from zero, that is, according to (3.11) the entropy $H_n$ change. And according to the proved theorems, non-conservation of the phase volumes $(\vec{r},\vec{v}),(\vec{r},\vec{v},\dot{\vec{v}}),...$ (but it corresponds to conservation of $(\vec{r},\vec{v},...),(\vec{v},\dot{\vec{v}},...),...$) takes place in general case, which leads to dissipative systems appearance.

In fact the generalized phase «liquid» in the infinite phase space $^{\Psi}\Omega'$ is «incompressible», according to Theorem 1, and the sources $Q_\infty = 0$ (3.14) and generalized Liouville's equation is fulfilled (3.8). Thus, a physical system in the infinite phase space $^{\Psi}\Omega'$ is a conservative one.

The transition to a lower dimensionality of the phase space $^{\Psi}_n\Omega'$, the phase «liquid» becomes «compressible» as new sources of entropy $Q_n \neq 0$ appear and generalized Liouville's equation (3.8) falls into infinite Vlasov equation chain (3.5) and infinite entropy equation chain (3.11) corresponding to it. Thus, a physical system becomes a dissipative one in general case.

At that, a special case of the second Vlasov equation in (3.5) is the known Liouville's equation. In the absence of the sources $Q_2 = 0$ from (3.5), (3.13) and (3.11), we obtain

$$\frac{\partial f_2}{\partial t} + f_2 \operatorname{div}_r \vec{v} + \vec{v}\nabla_r f_2 + f_2 \operatorname{div}_v \langle \dot{\vec{v}} \rangle + \left(\nabla_v f_2, \langle \dot{\vec{v}} \rangle\right) = 0,$$

$$\frac{\partial f_2}{\partial t} + \vec{v}\nabla_r f_2 + \left(\nabla_v f_2, \langle \dot{\vec{v}} \rangle\right) = -f_2 Q_2 = 0,$$

$$\frac{\partial f_2}{\partial t} + \vec{v}\nabla_r f_2 + \left(\nabla_v f_2, \langle \dot{\vec{v}} \rangle\right) = 0. \tag{3.16}$$

According to [6], in special case at $\langle \dot{\vec{v}} \rangle = \frac{d_1}{dt}\langle \vec{v} \rangle$ (for non-dissipative systems), equation (3.16) transforms into classical Liouville's equation. Note, that in general case (of dissipative systems), identical relation [6] takes place

$$\langle \dot{\vec{v}} \rangle = \frac{d_1}{dt}\langle \vec{v} \rangle + \left(\vec{v} - \langle \vec{v} \rangle\right)\left(\vec{v} - \langle \vec{v} \rangle, \nabla_r S_2\right) + \vec{\eta}, \tag{3.17}$$

where $\vec{\eta}$ is some vector-function, having this property

$$\int_{(\infty)} f_2 \vec{\eta} d^3 v = f_1 \langle \vec{\eta} \rangle = 0 \Rightarrow \langle \vec{\eta} \rangle = 0. \tag{3.18}$$

The value

$$\vec{F}_d = m\left(\vec{v} - \langle \vec{v} \rangle\right)\left(\vec{v} - \langle \vec{v} \rangle, \nabla_r S_2\right) + \vec{\eta}',$$

where $m$ has the dimensionality of mass, may be interpreted physically as a dissipative force.

**Conclusions**

Thus, the paper considers the infinite dimensional phase space $^{\Psi}\Omega'$ ($^{\Psi}_T\Omega'$). The evolution of the physical system in the space $^{\Psi}\Omega'$ develops along the generalized phase trajectories $\vec{\xi}(t)$. The generalized phase trajectories $\vec{\xi}(t)$ are analytical functions and never cut one another. The evolution of the system is defined by Taylor's mapping, which has a Jacobian matrix determinant set to unit (see Theorems 1-3). It's shown that the conservation of the classical phase volume is possible in the space $^{\Psi}_2\Omega''$ only. And in the space $^{\Psi}_2\Omega'$, the classical phase volume is not conserved (see Example §2).

The velocity vector field $\vec{u}(\vec{\xi})$ in the space $^{\Psi}\Omega'$ has no sources $Q_{\infty} = \text{div}_{\xi}\vec{u} = 0$. The absence of the sources $Q_{\infty} = 0$ leads to incompressibility of the phase «liquid» in the space $^{\Psi}\Omega'$. As a result, the probability density function $f_{\infty}$ satisfies generalized Liouville's equation (3.8) and the system's entropy is independent of time $H_{\infty} = const$. Generalized Liouville's equation (3.8) allows solution in the form of characteristics, along which $f_{\infty}(\vec{\xi}(t), t) = const$. At that, the characteristics are the generalized phase trajectories $\vec{\xi}(t)$ (1.1).

As a result, the physical system in the infinite dimensional space $^{\Psi}\Omega'$ is a conservative one.

At transition to finite phase spaces $^{\Psi}_n\Omega'$ by averaging (3.13) leads to appearance of dissipation sources $Q_n \neq 0$ in general case as the non-intersecting generalized phase trajectories $\vec{\xi}(t)$ «project» into finite dimensional spaces $^{\Psi}_n\Omega'$. The «projections» of the trajectories $\vec{\xi}(t)$ may have intersection points, in which the velocity divergence does not equal zero (3.12). As a result the phase «liquid» in the space $^{\Psi}_n\Omega'$ becomes compressible in general case.

Appearance of the sources $Q_n \neq 0$ leads to unsteadily of the entropies $H_n$ (3.11). Generalized Liouville's equation (3.8) falls into the Vlasov equations chain (3.5) for the functions $f_1(\vec{r},t), f_2(\vec{r},\vec{v},t), f_3(\vec{r},\vec{v},\dot{\vec{v}},t),...$ at transition to the finite spaces $^{\Psi}_n\Omega'$. Thus, the Vlasov equations chain (3.5) is an analogue of generalized Liouville's equation (3.8) in finite dimensional spaces $^{\Psi}_n\Omega'$.

In special case of a non-dissipative system, the modified Vlasov equation (3.16-17) [6] transforms into classical Liouville's equation for the function $f_2(\vec{r},\vec{v},t)$ and the classical phase volume is conserve in the space $^{\Psi}_n\Omega''$.

**Appendix**

*Proof of Theorem 1*
Let us write the Jacobian determinant for the Taylor mapping:

$$|J| = \begin{vmatrix} \dfrac{\partial x_2}{\partial x_1} & \dfrac{\partial x_2}{\partial v_1^{(x)}} & \dfrac{\partial x_2}{\partial \dot{v}_1^{(x)}} & \cdots \\ \dfrac{\partial y_2}{\partial x_1} & \dfrac{\partial y_2}{\partial v_1^{(x)}} & \dfrac{\partial y_2}{\partial \dot{v}_1^{(x)}} & \cdots \\ \dfrac{\partial z_2}{\partial x_1} & \dfrac{\partial z_2}{\partial v_1^{(x)}} & \dfrac{\partial z_2}{\partial \dot{v}_1^{(x)}} & \cdots \\ \dfrac{\partial v_2^{(x)}}{\partial x_1} & \dfrac{\partial v_2^{(x)}}{\partial v_1^{(x)}} & \dfrac{\partial v_2^{(x)}}{\partial \dot{v}_1^{(x)}} & \cdots \\ \cdots & \cdots & \cdots & \end{vmatrix} \quad (A.1)$$

The Taylor mapping has the form:

$$\vec{r}^{(2)} = \vec{r}^{(1)} + \vec{v}^{(1)}\tau + \frac{\dot{\vec{v}}^{(1)}}{2!}\tau^2 + \frac{\ddot{\vec{v}}^{(1)}}{3!}\tau^3 + ....$$

$$\vec{v}^{(2)} = \vec{v}^{(1)} + \dot{\vec{v}}^{(1)}\tau + \frac{\ddot{\vec{v}}^{(1)}}{2!}\tau^2 + \frac{\dddot{\vec{v}}^{(1)}}{3!}\tau^3 + .... \quad (A.2)$$

$$\dot{\vec{v}}^{(2)} = \dot{\vec{v}}^{(1)} + \ddot{\vec{v}}^{(1)}\tau + \frac{\dddot{\vec{v}}^{(1)}}{2!}\tau^2 + ....$$

.....

The quantity $\tau$ in (A.2) is a fixed parameter of the Taylor mapping. Substituting (A.2) in the determinant (A.1) and calculating it we obtain:

$$|J| = \begin{vmatrix} 1 & 0 & 0 & \tau & 0 & 0 & \frac{\tau^2}{2} & 0 & 0 & ... \\ 0 & 1 & 0 & 0 & \tau & 0 & 0 & \frac{\tau^2}{2} & 0 & ... \\ 0 & 0 & 1 & 0 & 0 & \tau & 0 & 0 & \frac{\tau^2}{2} & ... \\ 0 & 0 & 0 & 1 & 0 & 0 & \tau & 0 & 0 & ... \\ 0 & 0 & 0 & 0 & 1 & 0 & 0 & \tau & 0 & ... \\ ... & ... & ... & ... & ... & ... & ... & ... & .... & ... \\ 0 & ... & ... & ... & ... & ... & ... & .... & ...0 & 1 \end{vmatrix} = 1 \quad (A.3)$$

The result (A.3) proves Theorem 1.

*Proof of Theorem 2*

Writing the Jacobian matrix determinant for an X-projection Taylor mapping, we obtain:

$$\left|J^{(x)}\right| = \begin{vmatrix} \frac{\partial x_2}{\partial x_1} & \frac{\partial x_2}{\partial v_1^{(x)}} & \frac{\partial x_2}{\partial \dot{v}_1^{(x)}} & ... \\ \frac{\partial v_2^{(x)}}{\partial x_1} & \frac{\partial v_2^{(x)}}{\partial v_1^{(x)}} & \frac{\partial v_2^{(x)}}{\partial \dot{v}_1^{(x)}} & ... \\ \frac{\partial \dot{v}_2^{(x)}}{\partial x_1} & \frac{\partial \dot{v}_2^{(x)}}{\partial v_1^{(x)}} & \frac{\partial \dot{v}_2^{(x)}}{\partial \dot{v}_1^{(x)}} & ... \\ ... & ... & ... & ... \end{vmatrix} \quad (A.4)$$

X-projection of Taylor's mapping is of the form:

$$x_2 = x_1 + v_1^{(x)}\tau + \frac{\dot{v}_1^{(x)}}{2!}\tau^2 + \frac{\ddot{v}_1^{(x)}}{3!}\tau^3 + ....$$

$$v_2^{(x)} = v_1^{(x)} + \dot{v}_1^{(x)}\tau + \frac{\ddot{v}_1^{(x)}}{2!}\tau^2 + \frac{\dddot{v}_1^{(x)}}{3!}\tau^3 + ....$$ (A.5)

$$\dot{v}_2^{(x)} = \dot{v}_1^{(x)} + \ddot{v}_1^{(x)}\tau + \frac{\dddot{v}_1^{(x)}}{2!}\tau^2 + ....$$

.....

Note, that the value $\tau$ in expression (A.5) is a fixed parameter of mapping. Writing special derivatives, which stand in determinant (A.4), taking into account (A.5), we obtain:

$$\frac{\partial x_2}{\partial x_1} = 1, \quad \frac{\partial x_2}{\partial v_1^{(x)}} = \tau, \quad \frac{\partial x_2}{\partial \dot{v}_1^{(x)}} = \frac{\tau^2}{2!}, \quad \frac{\partial x_2}{\partial \ddot{v}_1^{(x)}} = \frac{\tau^3}{3!}, ....$$

$$\frac{\partial v_2^{(x)}}{\partial x_1} = 0, \quad \frac{\partial v_2^{(x)}}{\partial v_1^{(x)}} = 1, \quad \frac{\partial v_2^{(x)}}{\partial \dot{v}_1^{(x)}} = \tau, \quad \frac{\partial v_2^{(x)}}{\partial \ddot{v}_1^{(x)}} = \frac{\tau^2}{2!}, ....$$ (A.6)

$$\frac{\partial \dot{v}_2^{(x)}}{\partial x_1} = 0, \quad \frac{\partial \dot{v}_2^{(x)}}{\partial v_1^{(x)}} = 0, \quad \frac{\partial \dot{v}_2^{(x)}}{\partial \dot{v}_1^{(x)}} = 1, \quad \frac{\partial \dot{v}_2^{(x)}}{\partial \ddot{v}_1^{(x)}} = \tau, ....$$

.....

Inserting expressions (A.6) into determinant (A.4) and calculating it, we obtain:

$$\left|J^{(x)}\right| = \begin{vmatrix} 1 & \tau & \frac{\tau^2}{2!} & ... \\ 0 & 1 & \tau & ... \\ 0 & 0 & 1 & ... \\ ... & ... & ... & ... \\ 0 & ... & ..0 & 1 \end{vmatrix} = 1$$ (A.7)

Analogue result is obtained for $J^{(y)}$ and $J^{(z)}$, that is:

$$\left|J^{(x)}\right| = \left|J^{(y)}\right| = \left|J^{(z)}\right| = 1$$ (A.8)

Which was to be proved.

*Proof of Theorem 3*

Indeed, in case of (1.7), the Jacobian matrix determinant (A.1) is of the form:

$$|J| = \left|\frac{\partial\left(x_2, v_2^{(x)}, ..., y_2, v_2^{(y)}, ..., z_2, v_2^{(z)}, ...\right)}{\partial\left(x_1, v_1^{(x)}, ..., y_1, v_1^{(y)}, ..., z_1, v_1^{(z)}, ...\right)}\right| =$$

$$=\begin{vmatrix} \dfrac{\partial x_2}{\partial x_1} & \dfrac{\partial x_2}{\partial v_1^{(x)}} & \dfrac{\partial x_2}{\partial \dot{v}_1^{(x)}} & \cdots \\ \cdots & \cdots & \cdots & \cdots \\ \dfrac{\partial y_2}{\partial x_1} & \dfrac{\partial y_2}{\partial v_1^{(x)}} & \dfrac{\partial y_2}{\partial \dot{v}_1^{(x)}} & \cdots \\ \cdots & \cdots & \cdots & \\ \dfrac{\partial z_2}{\partial x_1} & \dfrac{\partial z_2}{\partial v_1^{(x)}} & \dfrac{\partial z_2}{\partial \dot{v}_1^{(x)}} & \cdots \\ \cdots & \cdots & \cdots & \end{vmatrix} = \begin{vmatrix} 1 & \tau & \cdots & 0 & \cdots \\ \cdots & \cdots & & \cdots & \cdots \\ 0 & \cdots & 1 & \tau & \cdots & 0 & \cdots \\ \cdots & \cdots & & \cdots & & \\ 0 & \cdots & \cdots & 1 & \tau & \cdots & 0 & \cdots \\ \cdots & \cdots & & \cdots & & \end{vmatrix} = \begin{vmatrix} J^{(x)} & 0 & 0 \\ 0 & J^{(y)} & 0 \\ 0 & 0 & J^{(z)} \end{vmatrix} = 1$$

Where expressions (A.5)-(A.6) are taken into account. Theorem 3 is proved.

*Proof of Theorem 4*

Let $\vec{\xi} \in {}^{\Psi}\Omega''$. Let us consider arbitrary $\delta$ - neighborhood of the point $\vec{\xi}$:

$$\vec{\xi} = \{x, y, z, v^x, v^y, v^z, \dot{v}^x, \dot{v}^y, \dot{v}^z, \ldots\}^T \in {}^{\Psi}\Omega'' \tag{A.9}$$

$$\vec{\xi} \pm \vec{\Delta}_1 = \{x \pm \delta^x, y, z, v^x, v^y, v^z, \dot{v}^x, \dot{v}^y, \dot{v}^z, \ldots\}^T \in {}^{\Psi}\Omega''$$
$$\vec{\xi} \pm \vec{\Delta}_2 = \{x, y \pm \delta^y, z, v^x, v^y, v^z, \dot{v}^x, \dot{v}^y, \dot{v}^z, \ldots\}^T \in {}^{\Psi}\Omega'' \tag{A.10}$$
$$\vec{\xi} \pm \vec{\Delta}_3 = \{x, y, z \pm \delta^z, v^x, v^y, v^z, \dot{v}^x, \dot{v}^y, \dot{v}^z, \ldots\}^T \in {}^{\Psi}\Omega''$$

$$\xi \pm \Delta_4 = \{x, y, z, v^x \pm \delta v^x, v^y, v^z, \dot{v}^x, \dot{v}^y, \dot{v}^z, \ldots\}^T \notin {}^{\Psi}\Omega''$$
$$\xi \pm \Delta_5 = \{x, y, z, v^x, v^y \pm \delta v^y, v^z, \dot{v}^x, \dot{v}^y, \dot{v}^z, \ldots\}^T \notin {}^{\Psi}\Omega'' \tag{A.11}$$
$$\xi \pm \Delta_6 = \{x, y, z, v^x, v^y, v^z \pm \delta v^z, \dot{v}^x, \dot{v}^y, \dot{v}^z, \ldots\}^T \notin {}^{\Psi}\Omega''$$
......

Six points in (A.11) do not belong to the subspace ${}^{\Psi}\Omega''$. Indeed, in (A.9) and in (A.10), the coordinates $x, y, z$ are used corresponding physically to the same point. However in expression (A.9) the velocity components $v^x, v^y, v^z$ correspond to the given point, and according to writing (A.11) the velocity components $v^x \pm \delta v^x, v^y, v^z$ or $v^x, v^y \pm \delta v^y, v^z$ or $v^x, v^y, v^z \pm \delta v^z$ correspond to the very same point. But the velocity, acceleration, high order accelerations are defined clearly for each kinematical point in the subspace ${}^{\Psi}\Omega''$, and in this case we have uncertainty in their values. The obtained contradiction points at impossibility of points of (A.11)-type points to belong to the subspace ${}^{\Psi}\Omega''$. Shifting $\pm\Delta$ succeeding components of the vector $\vec{\xi}$, we also obtain points not belonging to the subspace ${}^{\Psi}\Omega''$.

As a result, the maximum dimensionality of the subspace ${}^{\Psi}\Omega''$ equals three according to expressions (A.9-10), taking into account that the values $v^x, v^y, v^z, \dot{v}^x, \dot{v}^y, \dot{v}^z, \ldots$ are not independent for ${}^{\Psi}\Omega''$, and according to Definition 6 depend on the coordinates $x, y, z$. Theorem 6 is proved.

## Proof of Corollary 4

We execute proving only for the space $\omega_x$. By analogue with Theorem 6 proof, we write the coordinates of points (A.9)-(A.11) in the space $\omega_x$:

$$\vec{\eta}^{(x)} = \{x, v^x, \dot{v}^x, .....\}^T \in \omega_x, \quad (A.12)$$

$$\vec{\eta}^{(x)} \pm \vec{\Delta}_1 = \{x \pm \delta^x, v^x, \dot{v}^x, .....\}^T \in \omega_x, \quad (A.13)$$

$$\vec{\eta}^{(x)} \pm \vec{\Delta}_2 = \{x, v^x \pm \delta v^x, \dot{v}^x, .....\}^T \in \omega_x, \quad (A.14)$$

......

As a physical system may content an entire multiplicity of kinematical points with the same coordinates along the OX-axis, and, at that, different kinematical characteristics $v^x, \dot{v}^x, .....,$ then expressions (A.12)-(A.14) are non-contradictory, consequently, the dimensionality of the space $\omega_x$ may be greater than unity, which was to be proved.

## Proof of Theorem 5

Let us multiply the equation in (3.8) by $(1 + \ln f_\infty)$ and integrate it over $d^3\omega = d\omega_x d\omega_y d\omega_z$, we obtain [6, 20, 21]

$$(1+\ln f_\infty)\frac{\partial f_\infty}{\partial t} + (1+\ln f_\infty)\operatorname{div}_\xi[\vec{u}f_\infty] = 0,$$

$$\int_{(\infty)} \frac{\partial}{\partial t}(f_\infty \ln f_\infty) d^3\omega + \int_{(\infty)} (1+\ln f_\infty)\operatorname{div}_\xi[\vec{u}f_\infty] d^3\omega = 0. \quad (A.15)$$

The first integral in (A.15) is of the form:

$$\int_{(\infty)} \frac{\partial}{\partial t}(f_\infty \ln f_\infty) d^3\omega = -\frac{\partial}{\partial t}\int_{(\infty)} f_\infty S_\infty d^3\omega = -\frac{d}{dt}\left[N(t)\langle...\langle S_\infty\rangle...\rangle(t)\right]. \quad (A.16)$$

The second integral in (A.15)

$$\int_{(\infty)} (1+\ln f_\infty)\operatorname{div}_\xi[f_\infty \vec{u}] d^3\omega = \int_{(\infty)} (\vec{u},(1+\ln f_\infty)\nabla_\xi f_\infty) d^3\omega + \int_{(\infty)} (1+\ln f_\infty) f_\infty \operatorname{div}_\xi \vec{u}\, d^3\omega =$$

$$= \int_{(\infty)} (\vec{u},\nabla_\xi(f_\infty \ln f_\infty)) d^3\omega + \int_{(\infty)} f_\infty \operatorname{div}_\xi \vec{u}\, d^3\omega + \int_{(\infty)} f_\infty \ln f_\infty \operatorname{div}_\xi \vec{u}\, d^3\omega =$$

$$= -\int_{(\infty)} (\vec{u},\nabla_\xi(f_\infty S_\infty)) d^3\omega - \int_{(\infty)} f_\infty S_\infty \operatorname{div}_\xi \vec{u}\, d^3\omega + \int_{(\infty)} f_\infty \operatorname{div}_\xi \vec{u}\, d^3\omega =$$

$$= -\int_{(\infty)} \operatorname{div}_\xi[f_\infty S_\infty \vec{u}] d^3\omega + \int_{(\infty)} f_\infty \operatorname{div}_\xi \vec{u}\, d^3\omega = \int_{(\infty)} f_\infty Q_\infty d^3\omega = N(t)\langle...\langle Q_\infty\rangle...\rangle(t), \quad (A.17)$$

where the following condition is supposed to be fulfilled $\int_{(\infty)} \operatorname{div}_\xi[f_\infty S_\infty \vec{u}] d^3\omega = \int_{\Sigma_\infty} f_\infty S_\infty \vec{u} d\vec{\sigma} = 0$

(this condition is imposed on the functions $f_\infty$ as a default at deducing the chain of the Vlasov equations [20, 21, 27-29]).

Using (A.16) and (A.17) in (A.15), we obtain ultimately

$$\frac{d}{dt}\left[N(t)\langle...\langle S_\infty\rangle...\rangle(t)\right] = N(t)\langle...\langle Q_\infty\rangle...\rangle(t). \tag{A.18}$$

If the number of particles is constant, then $N(t) = N_0 = const$, and expression (A.19) is as follows

$$\frac{d\langle...\langle S_\infty\rangle...\rangle}{dt} = \langle...\langle Q_\infty\rangle...\rangle.$$